# From Quantitative Spatial Operator to Qualitative Spatial Relation Using Constructive Solid Geometry, Logic Rules and Optimized 9-IM Model

A Semantic Based Approach

Helmi Ben Hmida, Frank Boochs
Institut i3mainz, am Fachbereich Geoinformatik und Vermessung
Fachhochschule Mainz, Lucy-Hillebrand-Str. 2 55128
Mainz, Germany
e-mail: {helmi.benhmida, boochs}@geoinform.fh-mainz.de

Christophe Cruz, Christophe Nicolle
Laboratoire Le2i, UFR Sciences et Techniques

Université de Bourgogne
B.P. 47870, 21078 Dijon Cedex, France
e-mail: {christophe.cruz, cnicolle}@u-bourgogne.fr

*Abstract*—The Constructive Solid Geometry (CSG) is a data model providing a set of binary Boolean operators such as Union, Difference and Intersection. In this work, these operators are used to compute topological relations between objects defined by the constraints of the nine Intersection Model (9-IM) from Egenhofer. With the help of these constraints, we define a procedure to compute the topological relations on CSG objects. These topological relations are Disjoint, Contains, Inside, Covers, CoveredBy, Equals and Overlaps, and are defined in a top-level ontology with a specific semantic definition on relation such as Transitive, Symmetric, Asymmetric, Functional, Reflexive, and Irreflexive. The results of topological relations computation are stored in the ontology allowing after what to infer on these topological relationships. In addition, logic rules based on the Semantic Web Language allows the definition of logic programs that define which topological relationships have to be computed on which kind of objects. For instance, a "Building" that overlaps a "Railway" is a "RailStation".

## I. INTRODUCTION

"Qualitative spatial relations are symbols abstraction of geometric representation, which allow computational analyses independent of, but consist with, graphical depiction" [1]. Qualitative spatial relationships are used in many areas of Computer Science. Actually, reasoning about such relationships is fundamental to infer about graphical depiction through logic mechanisms. In addition, these relationships facilitate the access to data by a query processing mechanism that refers to objects and their relationships. Qualitative spatial reasoning is appropriate for prediction and diagnoses of physical systems in a qualitative manner, especially when no quantitative description is available or computationally intractable. Methods for modelling spatial relationships have been compiled in several surveys such as [3][4]. Current models for topological relationships belongs to two main categories – connection based [4], and intersection based [5]. The both models fall to the same topological relationships for the two simple 2D regions. From a logical point of view, the qualitative models are defined to infer on topological relations without taking into account real geometries. Operators were defined on these relationships allowing the specification of a spatial query language. The Open Geospatial Consortium (OGC) defined a standard nomination to the basic topological relations [6]. Topological operators are used to query the topological relationship between two spatial entities. These relations and operators are between intervals in $\mathbb{R}^1$ and for regions in $\mathbb{R}^2$. Zlatanova in [7][18] gives a survey on different 3D models and relations in $\mathbb{R}^3$. The spatial operators available for spatial query language consist of 3D Topological operators (disjoint, within, contains, etc.) [8], 3D Metric operators (distance, closerThan, fartherThan, etc.) [9], 3D Directional operators (above, below, northOf, etc.) [10] and finally 3D Boolean operators (union, intersection, etc.) [11].

From the $\mathbb{R}^3$ space implementation point of view in [9], the octree-based implementation [13] and the B-Rep approaches are used to define the spatial operators of a query language [12]. In the octree-approach, octrees allows the application of recursive algorithms that successively increase the discrete resolution of the spatial objects employed. The B-Rep, approach is used for metric operators such as *mindist, maxdist, isCloserto* and *isFartherfrom relations.* The bounding facets of each operand are indexed by a, so-called, axis-aligned bounding boxes tree (AABB tree). The algorithm uses the AABB-tree structure to identify candidate pairs of facets, for which an exact, but expensive distance algorithm is employed.

From the semantics point of view, the qualitative spatial relations are used to perform inference and to identify inconsistencies on these relations. An ontology based approach is described in [14] and focuses on regions in $\mathbb{R}^2$.

In our current work, we will focus more on the $\mathbb{R}^3$ dimension environment where the 3D topologic relation computation is carried out by external libraries which made the execution process more optimal. The presented approach aims at defining topological relations based on the optimized 9 Intersection Model in $\mathbb{R}^3$ [2], and compute them with the Boolean operators defined by Constructive Solid Geometry (CSG) [21]. Actually, the 9-IM model is widely used to represent spatial relations in $\mathbb{R}^2$. These relations exist also in $\mathbb{R}^3$ with much more variation and complexity. In the actual contribution, the quantitative spatial operators are implemented using built-ins based the Semantic Web Rules Languages (SWRL) which allows the definition of logic program base on Horn-like clauses. This language is designed to perform logical program on Ontology Web Language (OWL). Consequently, the results of these 3D spatial operators may enrich the ontology with spatial relations between the different object represented via CSG model.

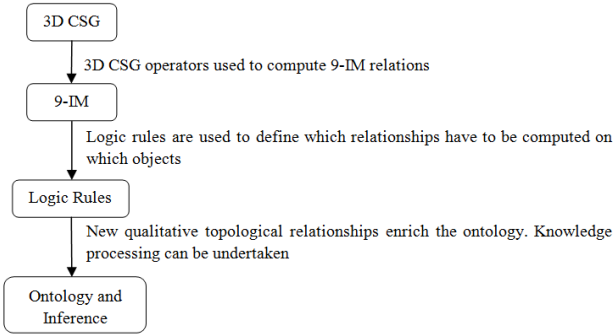

Figure 1. General overview of the process sequence from ontology to CSG geometric data model.

Figure 1 depicts the process sequence for the enrichment of an ontology containing 3D objects. The ontology is populated with topological relationships based on SWRL rules. Once done, the inference process on these relationships makes a step forward to infer new knowledge out. The used logic rules are based on new topological built-ins defined based on the 9-IM constraints, and computed for each object´s relation using 3D CSG objects and Boolean operators. The following example is a SWRL rule that uses the "topo:overlaps" built-ins.

*Building(?b) ^ Railway(?r) ^ topo: overlaps(?b, ?r) → RailStation(?b)*

This paper is divided into 5 sections. Section 2 introduces the technical background on 9-IM, Constructive Solid Geometry operators, and logic rules. Section 3 deals with the important elements of the topologic qualification process implementation. Section 4 shows the first results and section 5 concludes this paper.

## II. BACKGROUND

Spatial object representation can be divided into two main categories, the surface-based representation and the volume-based representation [19]. The surface-based representations are materialized by the grid model, the shape model, the facet model and the boundary representation (B-Rep). The volume-based representations are taking place via the 3D array, the octree, the constructive solid geometry (CSG) and the 3D TIN. Such a spatial object representations are used to compute spatial and topological relations by the use of unary or binary operators. Among the different defined spatial relationship, we put the light in this paper on the 3D topological operators.

### A. Topological relationships

Spatial reasoning is a process that uses spatial theory and artificial intelligence to model and to analyze spatial relations between objects. The standard models are composed by the Simple Feature Relations, the Egenhofer Relations and the RCC8 Relations [17]. The Simple Features Relations are based on the defined standard of OGC and are composed of the following relationships: Equals, Disjoint, Intersects, Touches, Within, Contains, Overlaps, and Crosses [16]. The Egenhofer Relations [15] are composed of the following relationships: Equals, Disjoint, Overlap, Covers, Covered by, Inside, Contains [1]. Finally, the RCC8 Relations are presented by: Equals, Disconnected, Externally connected, Partially overlapping, Tangential proper part inverse, Tangential proper part, Non-tangential proper part, Non-tangential proper part inverse [17]. Initially, binary topological relations between two objects, A and B, are defined in terms of the four intersections of A's boundary (δA) and interior (A°) with the boundary (δB) and interior (B°) of B. Recently, this model has been extended by considering the location of each interior and boundary with respect to the other object's exterior. Therefore, the binary topological relation between two objects A and B, in $\mathbb{R}^2$ is based upon the intersection of A's interior (A°), boundary (δA), and exterior (A-) with B's interior (B°), boundary (δB), and exterior (B-). The 9 intersections between the six object´s parts describe a topological relation and can be concisely represented by a 3x3 matrix, called the 9-Intersection Model. The binary relationship R(A,B) between the two objects is then identified by composing all the possible set intersections of the six topological primitives, i.e. A°∩B°, δA∩B°, A- ∩ B°, A°∩δB, δA∩δB, A-∩δB, A°∩ B-, δA∩B-, A- ∩ B-, and qualifying empty (∅) or non-empty (¬∅) intersections. For example, if two objects have a common boundary, the intersection between the boundaries is non-empty, i.e. δA∩δB = ¬∅. If they have intersecting interiors, then the intersection A°∩ B° is not empty, i.e. A°∩ B° = ¬∅, Table 1.

TABLE I.  THE 9-IM MATRIX

$$R_{(A,B)} = \begin{pmatrix} A° \cap B° & A° \cap \delta B & A° \cap B^- \\ \delta A \cap B° & \delta A \cap \delta B & \delta A \cap B^- \\ A^- \cap B° & A^- \cap \delta B & A^- \cap B^- \end{pmatrix}$$

Actually, a spatial region (BIM or GIS) has simply three topologically distinct parts: the interior, the boundary, and the exterior. Specifying any part of the first geometry will completely determines the region of the other parts. Based on this observation, it appears reasonable to assume that topological relations between regions can be characterized by considering the intersections of any pair of parts, for example, boundary/exterior or interior/exterior, rather than only the boundary/interior intersections. To assess such alternatives, one has to determine whether the 4-intersection based on the boundary / interior / intersections is equivalent to one based on boundary / exterior or interior / exterior intersections. If so, the characterization of topological relations would have to be the same in each case. Based on this assumption, we opt to use the 9-IM principle in more optimal way by reducing it to a four intersection model based on the interior / exterior of each 3D geometry, Table2.

TABLE II. OPTIMIZED 9IM MODEL (LEFT) AND THE CORRESPONDENT GRAPHICAL REPRESENTATION (RIGHT)

$$R_{(A,B)} = \begin{pmatrix} A° \cap B° & A° \cap B^- \\ A^- \cap B° & A^- \cap B^- \end{pmatrix}$$

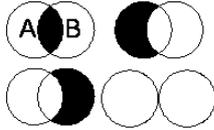

### B. CSG-based implementation of 3D topological relations

The Constructive Solid Geometry is defined by Friedrich A. Lohmüller in [21]. It is a technique used in solid modelling. It allows the modeller to create a complex surface or object by using Boolean operators such as union, intersection or difference to combine objects. A CSG object can be represented by a tree, where each leaf represents a primitive and each node a Boolean operation. There are only five CSG standard defined operations, which are materialized by the union, the intersection, the difference, the inverse and Clipped_by. These methods return the resulting solid of the operation and are restricted to objects including a closed space. Actually, lines and planes are both objects which do not enclose a volume, as consequence, no possible CSG operations can be applied on them. As a conventional solution, a solid will be created from a line and plane by adding a small noise rate to the above mentioned geometry, always with respect to the fact that the added noise rate is always less than those related to the used instrument during the survey of real objects. In the next, we will mainly focus on the implication of the CSG topologic operator within the optimized 9-IM model to qualify directly 3D topologic relation.

### C. Ontology and rules

The term Ontology has been used for centauries to define an object philosophically. Within the computer science domain, ontology is a formal representation of the knowledge through the hierarchy of concepts and the relationships between those concepts. In theory, ontology is a formal, explicit specification of shared conceptualization [24]. In any case, ontology can be considered as formalization of knowledge representation and Description Logics (DLs) and provides a logical formalization to the Ontologies [25].

Description logics (DLs) [26][27] are a family of knowledge representation languages that can be used to represent knowledge of an application domain in a structured and formally well-understood way. The following example defines a Mother as a Woman which has at least a child type of Person. By inference, it means that every individual type of Women which as at least a relation with a Person and the type of the relation is "hasChild", then this Woman is of kind of Mother.

$$Mother \equiv Woman \sqcap \exists hasChild.Person$$

As the Semantic Web technologies matured, the need of incorporating the concepts behind description logic within the ontology languages was realized. It took few generations for the ontology languages defined within Web environment to implement the description language completely. The Web Ontology Language (OWL) [28] is intended to be used when the information contained in documents needs to be processed by applications and not by human [29]. The horn logic, more commonly known the Horn clauses has been used as the base of logic programming and Prolog languages [31] for years. These languages allow the description of knowledge with predicates. The Horn logic has given a platform to define Horn-like rules through sub-languages of RuleML [23]. Summarizing, it could be said that ontology defines the data structure of a knowledge base and this knowledge base could be inferred through various inference engines. They can be performed under Horn logic through Horn-like rules languages. The following Horn-like rule is specified with the help of the SWRL language used to define rules. It means any parent of a child and the parent has a brother, then the brother is an uncle.

*Parent(?p, ?c) ^ Brother(?p, ?b) → Uncle(?b, ?c)*

The set of built-ins for SWRL is motivated by a modular approach that will allow further extensions in future releases within a (hierarchical) taxonomy. SWRL's built-ins approach is also based on the reuse of existing built-ins in XQuery and XPath, which are themselves based on XML Schema by using the Data types. These built-ins are keys for any external integration, like the integration of the topological operators. Built-ins in the SWRL can be used with the standard SWRL expressions. The built-ins process the rule expressions to deduce the result and couple with the

standard expression to return the results. For example

*Person(?x) ^ hasHeight(?x, ?h) ^ swrlb:greaterThan(?h, 6.5) → Tall(?x)*

### D. Enrichment of the ontology from Boolean operators using CSG model

The use of CSG model and its associated Boolean operator allows us to model the topological relationships. In order to combine SWRL rules with topological operators, news built-ins are defined to compute the operator. Consequently, the results of the operators can be used to define queries or enrich the ontology with new topological relationships between two objects. The following rule specifies that a "Building" defined in the ontology that overlaps a "Railway" defined as well in the ontology, is a "RailStation".

*Building(?b) ^ Railway(?r) ^ topo: overlaps(?b, ?r) → RailStation(?b)*

To make it realistic, two issues appear and have to be solved. First, the semantic definition of the relationships has to be defined in the ontology level regarding their own properties. Second, the calculation of topological relationships using Boolean operators has to be defined regarding the constraints of the optimized 9-IM model.

### III. IMPLEMENTATION

This section is divided into three sub-sections. The first one describes how Boolean operators are used to compute the optimized 9-IM matrix for a topological relation. The second introduces news relationships in the top-level ontology and its built-in. Finally, the last section deals with the translation engine which allows the computation of the topological built-ins to enrich the ontology.

### A. Calculation of 9-IM topological relationships using the CSG Boolean operators

Regarding the optimized 9-IM matrix, Table 2, only operators about intersection (A ∩ B), interior (A° equivalent A), complement ($A^-$ is equivalent to $\bar{A}$) are necessary. Table 3 presents relatively the new suggested mask for 3D topologic operations based on the interior and the exterior of each solid geometry. In parallel, it presents the relative CSG operation corresponding to each part of the mask.

TABLE III. OPTIMIZED 9IM MODEL (ON THE LEFT) WITH THE EQUIVALENT MASK USING CSG OPERATORS (ON THE RIGHT)

$$R_{(A,B)} = \begin{pmatrix} A° \cap B° & A° \cap B^- \\ A^- \cap B° & A^- \cap B^- \end{pmatrix} \quad R_{(A,B)} = \begin{pmatrix} A \cap B & A \backslash B \\ B \backslash A & \bar{A} \cap \bar{B} \end{pmatrix}$$

Table 4 shows the optimized 9-IM matrices of the topological predicates defined by Egenhofer.

TABLE IV. THE OPTIMIZED 9-IM MATRIX

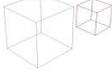

| $\begin{pmatrix} 0 & 1 \\ 1 & 1 \end{pmatrix}$ | $\begin{pmatrix} 1 & 1 \\ 0 & 1 \end{pmatrix}$ | $\begin{pmatrix} 1 & 1 \\ 1 & 1 \end{pmatrix}$ |
|---|---|---|
| A disjoint B  B disjoint A | A contains B  B inside A | A overlaps B  B overlaps A |

Table 5 presents the equivalent qualitative relations for each CSG operator. If one of these equations is false, the relation between the two objects cannot be verified.

TABLE V. EQUIVALENT QUALITATIVE RELATIONS TO QUALITATIVE CSG OPERATOR

| spatial relation | CSG operators |
|---|---|
| Disjoint | $(A \cap B = \emptyset) \land (A \backslash B = \neg \emptyset) \land (B \backslash A = \neg \emptyset) \land (\bar{A} \cap \bar{B} = \neg \emptyset)$ |
| Contain | $(A \cap B = \neg \emptyset) \land (A \backslash B = \neg \emptyset) \land (B \backslash A = \emptyset) \land (\bar{A} \cap \bar{B} = \neg \emptyset)$ |
| Overlaps | $(A \cap B = \neg \emptyset) \land (A \backslash B = \neg \emptyset) \land (B \backslash A = \neg \emptyset) \land (\bar{A} \cap \bar{B} = \neg \emptyset)$ |

### B. Definition of topological relationships in the ontology and new built-ins

Regarding the ontology, the top level ontology is created to model the topological relationships. This ontology is used to enrich an existing knowledge base to make it possible to define topological relationships between objects. The next table summarizes for each topological relation, its name in the ontology using the prefix "swrl_topo", its semantic characteristics and the new built-in to automatize the computation of relations with the help of SWRL rules. The swrl_topo:inside relation is the inverse relation of swrl_topo:contains, and the relation swrl_topo:covers is the inverse relation of swrl_topo:coveredBy.

TABLE VI. DEFINITION OF THE TOPOLOGICAL RELATIONSHIPS AND ITS SEMANTICS

| Topological relationships | Property Name | Characteristics | SWRL built-ins |
|---|---|---|---|
| Disjoint | topo:disjoint | Transitive, Symmetric Irreflexive | swrl_topo:disjoint (?x, ?y) |
| Contains | topo:contains | Transitive, Asymmetric Irreflexive | swrl_topo:contains (?x, ?y) |
| Overlaps | topo:overlaps | Symmetric irreflexive | swrl_topo:overlaps (?x, ?y) |

### C. Translation engine

The translation engine allows the computation of spatial SWRL rules which can also be queries. It interprets the statements in order to parse the spatial components. Once parsed, they are computed through relevant spatial functions and operations by the translation engine through the operations provided at the CSG level. The results are populated in the knowledge base, thus making it spatially rich, Figure 2.

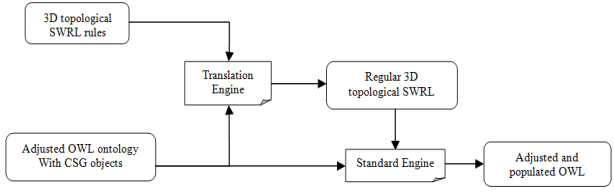

Figure 2. The translation engine that process 3D topologic SWRL rules with topological built-ins.

## IV. QUALITATIVE 3D TOPOLOGICAL RELATION EMPLOYMENT, TESTS AND RESULTS

Let´s first remind that our solution is based on a knowledge base structure instead of any standard data bases to improve the portability, the sharing degree of the document and mostly to add a new semantic dimension to the acquired data. Figure 3 shares a print screen of the qualified 3D topological relationships between geometries detected from an airport scene. In the next, two main kind of spatial queries on the knowledge base with a strongly formalized foundation to maximize the exploitation of the implemented semantic 3D topological relations will be highlighted.

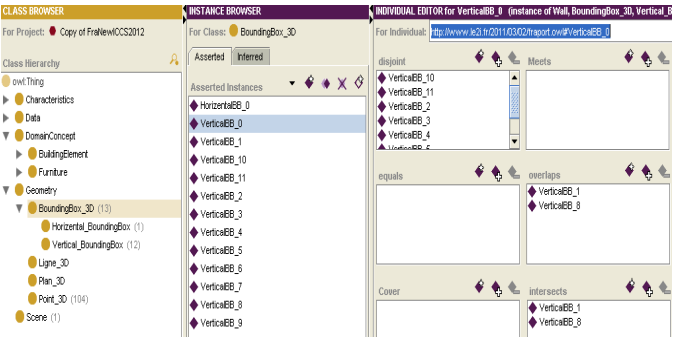

Figure 3. Qualified 3D topologic relations between an ontology individuals

### A. SWRL based use case

Known as SWRL, Semantic Web Rules Languages with the extended built-Ins to support the 3D topological processing is performed. As seen in mentioned SWL rule, the antecedent part is composed by classes like Wall in this case, data properties, but also built-Ins for 3D topology that will be later on converted to simple object properties in this case. In the consequent part, and once the topological assertion is verified, overlaps in this case, the (?y) elements will be classified as a semantic element from the class Wall.

$Wall(?x) \wedge VerticalBoundingBox(?y) \wedge swrl\_topo:overlaps(?x,?y) \wedge hasheight(?y,?h) \wedge swrlb:greaterThan(?h,3) \rightarrow Wall(?y)$

### B. SQWRL based uses cases

As we have already selected a qualitative manner based on semantic knowledge to define spatial operators, SQWRL (**S**emantic **Q**uery-Enhanced **W**eb **R**ule **L**anguage) language can be used as a query language to query the knowledge base. The next equation is an example of a query that selections all distinct overlapping bounding box in the current knowledge base.

$Vertical\_BoundingBox(?x) \wedge Vertical\_BoundingBox(?y) \wedge swrl\_topo:overlaps(?x, ?y) \rightarrow sqwrl:selectDistinct(?x,?y)$

To test the performance of the quantitative 3D topological operators, a various number of geometries were created. The geometries can be much more complex but has to be closed without whole. During the tests, the query execution time is stored, Figure 4. Fewer than 500 boxes, the computation time stay almost constant. But upper than 500 boxes, the computation time is asymptotic. This is explained by the fact that the query process computes the Cartesian product of all the geometrical objects. Consequently, for k=2 and n=1000, the number of operations is almost an half million, and for n=10000 geometries the number of operations is almost 50 million. Consequently, after the computation of all relations, the query can be done only on the knowledge base without Boolean operators.

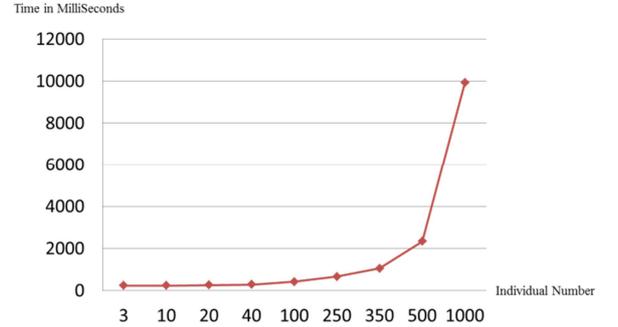

Figure 4. The SQWRL test performance up to 1000 boxes.

## V. CONCLUSION

This paper has presented a method to compute automatically topological relations using SWRL rules. The calculation of these rules is based on the definition of Constructive Solid Geometry. Some simplification will be undertaken regarding the 9-IM computation of each topological relationship in order to reduce the calculation volume. Future work on topological relationships will be undertaken also about basic rules that can be defined from [32] and depicted in the next SWRL rule. This can also be done by a composition of relations, meet ∘ contains ⊑ Disjoint.

$meet\ (?a, ?b) \wedge contains(?a, ?c) \rightarrow disjoint\ (?a, ?c)$